\documentclass[journal=aamick,manuscript=article]{achemso}
\usepackage[utf8]{inputenc}
\usepackage{graphicx}
\usepackage{chemfig}
\usepackage{siunitx}
\usepackage{dcolumn}
\usepackage{bm}
\usepackage{booktabs}
\usepackage[version=4]{mhchem}

\DeclareSIUnit{\atom}{atom}
\DeclareSIUnit{\elementarycharge}{\text{\ensuremath{e}}}

\title{Exploring Intrinsic and Extrinsic $p$-type Dopability of Atomically Thin $\beta$-\ce{TeO2} from First Principles}

\author{Rafael Costa-Amaral}
\affiliation{Institute for Materials Research, Tohoku University, Sendai 980-8577, Japan.}

\author{Soungmin Bae}
\affiliation{Institute for Materials Research, Tohoku University, Sendai 980-8577, Japan.}

\author{Vu Thi Ngoc Huyen}
\affiliation{Institute for Materials Research, Tohoku University, Sendai 980-8577, Japan.}

\author{Yu Kumagai}
\email{yukumagai@tohoku.ac.jp}
\affiliation{Institute for Materials Research, Tohoku University, Sendai 980-8577, Japan.}
\affiliation{Organization for Advanced Studies, Tohoku University, Sendai 980-8577, Japan}

\date{\today} 

\begin{document}

\begin{abstract}
Two-dimensional (2D) $\beta$-\ce{TeO2} has gained attention as a promising material
for optoelectronic and power device applications, thanks to its transparency and high hole mobility.
However, the underlying mechanism behind its $p$-type conductivity and dopability remains unclear.
In this study, we investigate the intrinsic and extrinsic point defects in monolayer and bilayer $\beta$-\ce{TeO2},
the latter of which has been experimentally synthesized,
using the HSE+D3 hybrid functional.
Our results reveal that most intrinsic defects are unlikely to contribute to $p$-type doping
in 2D $\beta$-\ce{TeO2}.
Moreover, \ce{Si} contamination could further impair $p$-type conductivity.
Since the point defects do not contribute to $p$-type conductivity,
we propose two possible mechanisms for hole conduction:
hopping conduction via localized impurity states, and substrate effects.
We also explored substitutional $p$-type doping in 2D $\beta$-\ce{TeO2} with 10 trivalent elements.
Among these, the \ce{Bi} dopant is found to exhibit a relatively shallow acceptor transition level.
However, most dopants tend to introduce deep localized states, where hole polarons
become trapped at \ce{Te}'s lone pairs.
Interestingly, monolayer $\beta$-\ce{TeO2} shows potential advantages over bilayers
due to reduced self-compensation effects for $p$-type dopants.
These findings provide valuable insights into defect engineering strategies for future electronic
applications involving 2D $\beta$-\ce{TeO2}.
\end{abstract}

\maketitle

\section{Introduction}\label{sec:introduction}
Wide band gap semiconductors have been extensively used in high-frequency and high-voltage applications,
serving as fundamental components for ultraviolet light emitters, power electronics, optoelectronics,
and radio-frequency devices~\cite{Yoshikawa_1_2007,Shi_2006230_2021}.
The fabrication of $n$-type oxide semiconductors has achieved remarkable success
with high electron mobility and efficiency,
such as in \ce{Al}-doped \ce{ZnO}~\cite{Majumder_16_2003}, \ce{Sn}-doped \ce{In2O3}~\cite{Kim_2045_2020},
and amorphous IGZO~\cite{Troughton_12388_2019}.
However, the lack of $p$-type counterparts with comparable performance limits the potential
for complementary metal-oxide-semiconductor technology and bipolar junction transistors,
essential components for high-efficiency power devices and integrated circuits,
although numerous computational efforts have been made to identify superior
transparent conducting oxides~\cite{Hautier.2013,PhysRevApplied.19.034063,gake2021point}.

Two-dimensional (2D) $\beta$-\ce{TeO2} has captured significant attention
as a potential high-mobility $p$-type semiconductor.
Calculations based on the density-functional theory first predicted high hole mobility of $\beta$-\ce{TeO2}~\cite{Guo_8397_2018}.
Zavabeti \textit{et al.} subsequently synthesized atomically thin
$\beta$-\ce{TeO2} on \ce{SiO2} substrates~\cite{Zavabeti_277_2021}.
These nanosheets exhibit a direct band gap of \SI{3.66}{\electronvolt}
and high hole mobility at room temperature,
which provides new possibilities for device applications, especially for field-effect transistors (FETs).
These results have attracted increasing attention to tellurium oxides,
with several studies reporting experimental and theoretical observations
of \ce{TeO2}~\cite{Shi_101901_2023,Yuan_305301_2024,Biswas_111343_2021,Dong_154382_2022,Guo_064010_2022,Liu_798_2024}.

While the $p$-type conductivity in 2D $\beta$-\ce{TeO2} is observed even without external doping,
the main origin of the $p$-type conductivity in 2D $\beta$-\ce{TeO2} still remains unexplored.
Point defects such as intrinsic vacancies and unintentional dopants are known sources of
hole carriers in semiconductors~\cite{Freysoldt_253_2014,Oba_060101_2018}.
Although they are expected to play crucial roles in hole conduction in 2D $\beta$-\ce{TeO2},
this hypothesis has not been verified experimentally and theoretically.
Previously, we have reported on the calculations of point defects
and impurities in the \textit{bulk} $\beta$-\ce{TeO2}~\cite{bulk_teo2_2024},
and found that no intrinsic defects contribute to hole conductivity.
This raises the question whether they would contribute to the $p$-type conductivity when the dimensionality is reduced.

Thus, our aim is to uncover the origins of $p$-type conductivity in atomically thin $\beta$-\ce{TeO2}
by calculating point defects, including unintentional dopants, using a
hybrid functional that reproduces the experimental band gap.
We primarily target the bilayer $\beta$-\ce{TeO2} because the synthesized 2D $\beta$-\ce{TeO2} is
composed of the bilayer in experiments~\cite{Zavabeti_277_2021}.
Additionally, we also focus on the monolayer to investigate the thickness effect.
As a result, we reveal that neither native defects nor \ce{Se} and
\ce{Si} impurities contribute to meaningful hole concentration.
To explain the experimentally observed $p$-type conductivity, we also discuss
the substrate effect by demonstrating charge transfer from quartz substrate to 2D $\beta$-\ce{TeO2}.
Finally, for applications such as photodetectors and solar cells, where controlling hole
concentration is crucial, we investigate potential acceptor dopants from 10 elements
capable of adopting a trivalent oxidation state at the Te sites.

\begin{figure}[t!]
\centering\includegraphics[width=0.5\linewidth]{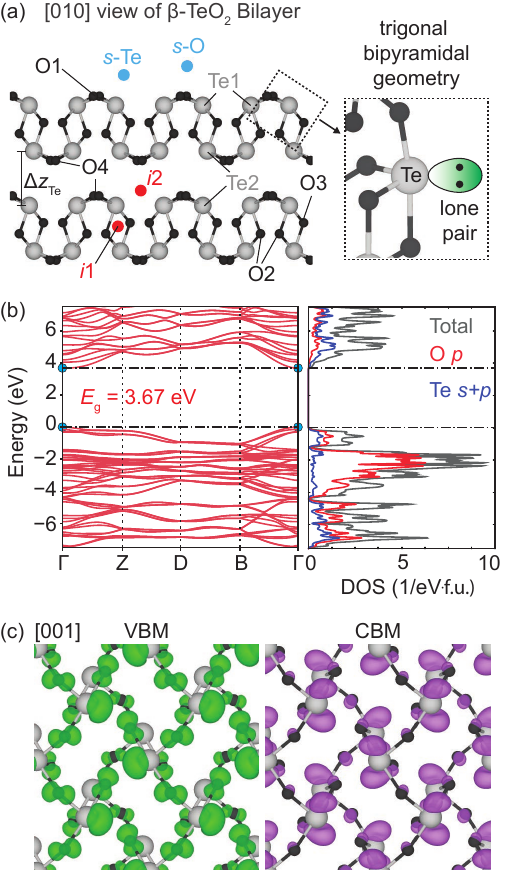}
\caption{(a) Side view of the optimized ${\beta}$-\ce{TeO2} bilayer, where the \ce{Te} and \ce{O} atoms are gray and black, respectively.
The nonequivalent \ce{Te} and \ce{O} sites are labeled accordingly along with the surface sites, $s$-\ce{Te},
    and $s$-\ce{O}, in blue, and the interstitial sites, $i1$, and $i2$, in red.
    A close up image of the structure depicts the \ce{Te} coordinated to four \ce{O} atoms
    in a distorted trigonal bipyramidal geometry.
    The schematic of the Te-5$sp$ lone pair orbital oriented toward the vacant space is also illustrated.
    (b) The band structure and density of states of the ${\beta}$-\ce{TeO2} bilayer calculated with the HSE+D3 functional.
    The energy along the $y$-axis is shown relative to the VBM.
    The dash-dotted lines represent the energy zero and the CBM,
    while the circles mark the VBM and CBM positions that are located at the $\Gamma$ point.
    (c) The partial charge densities of the VBM and CBM in the ${\beta}$-\ce{TeO2} bilayer viewed
    from the [001] direction. The isosurface levels are set to \SI{0.0005}{\elementarycharge\per\cubic{Bohr}}.}
\label{fig:perfect_TeO2}
\end{figure}

\section{Results and Discussion}\label{sec:results-and-discussion}
\subsection{Unit-cell parameters}\label{subsec:unit-cell-calculations}

Prior to studying point defects in  2D \ce{TeO2},
we benchmarked several exchange-correlation functionals including PBE~\cite{Perdew_3865_1996},
PBEsol~\cite{Perdew_136406_2008}, SCAN~\cite{Sun_036402_2015}, SCAN0~\cite{Hui_044114_2016}, PBE0~\cite{Adamo_6158_1999},
HSE06~\cite{Heyd_8207_2003,Krukau_224106_2006}, HSEsol~\cite{Schimka_024116_2011},
and HSE06 with the D3 Van der Waals correction~\cite{Grimme_154104_2010} (HSE+D3),
in predicting the structural properties and the band gap of the bilayer ${\beta}$-\ce{TeO2}.
The in-plane lattice constants and their ratio ($a$ and $b$, $b$/$a$), the \ce{Te}-\ce{Te} interlayer distance,
${\Delta}z_{\ce{Te}}$ (see Fig.~\ref{fig:perfect_TeO2}(a)), and the band gap $E_g$ are summarized in Table~\ref{tab:benchmark}.
While the PBE functional overestimates both in-plane lattice parameters, PBEsol markedly underestimates $a$, leading to a larger $b$/$a$ ratio.
As anticipated, both functionals underestimate the experimental band gap of \SI{3.66}{\electronvolt}
measured by Scanning Tunneling Spectroscopy (STS) on the nanosheet~\cite{Zavabeti_277_2021}.
The SCAN functional yields better results than PBE and PBEsol but still underestimates the band gap by \SI{1}{\electronvolt}.
Screened HSE functionals (HSE06 and HSEsol), predict the band gap to be, respectively, \SI{3.74}{\electronvolt} and \SI{3.51}{\electronvolt},
much closer to the experiment~\cite{Zavabeti_277_2021}.
The addition of dispersion corrections~\cite{Grimme_154104_2010} with the HSE06 functional (HSE+D3)
reduces the interlayer spacing from \SI{3.44}{\angstrom} to \SI{3.34}{\angstrom},
and yields the $b$/$a$ ratio and band gap (\SI{3.67}{\electronvolt}) closer to the experimental values.
On the other hand, unscreened hybrid functionals (PBE0 and SCAN0) largely overestimates
the band gap of the ${\beta}$-\ce{TeO2} bilayer, giving values of \SI{4.44}{\electronvolt} and \SI{5.37}{\electronvolt}.
Consequently, the HSE+D3 functional was selected to investigate the point defect properties in 2D ${\beta}$-\ce{TeO2}, as it most accurately reproduces experimental parameters.

\begin{table}
\centering\caption{In-plane lattice constants, $a$ and $b$, the $b$/$a$ ratio, the \ce{Te}-\ce{Te} interlayer distance,
${\Delta}z_{\ce{Te}}$ shown in Figure \ref{fig:perfect_TeO2}, and the band gap, $E_g$, of the ${\beta}$-\ce{TeO2} bilayer obtained from different exchange-correlation functionals.
    For both $a$ and $b$, the relative errors are also provided.
    The HSE+D3 functional is shown in \textbf{bold} as it is adopted for the point-defect calculations.}
\label{tab:benchmark}
\begin{tabular}{lcccccccc} \toprule
 xc functional      & $a$ ({\AA})  & Error (\%)    & $b$ ({\AA})  & Error (\%)   & $b$/$a$         & ${\Delta}z_{\ce{Te}}$ ({\AA}) & $E_g$ (\SI{}{\electronvolt}) \\ \bottomrule
PBE    & \tablenum{5.55} & \tablenum{1.54} & \tablenum{5.74} &  \tablenum{2.37} &\tablenum{1.04} & \tablenum{3.34}               & \tablenum{2.46}  \\
PBEsol & \tablenum{5.23} & \tablenum{-4.19} & \tablenum{5.69} & \tablenum{1.46} & \tablenum{1.09} & \tablenum{3.14}               & \tablenum{2.23}  \\
SCAN   & \tablenum{5.34} & \tablenum{-2.27} & \tablenum{5.62} & \tablenum{0.18} & \tablenum{1.05} & \tablenum{3.30}               & \tablenum{2.68}  \\
HSE06    & \tablenum{5.50} & \tablenum{0.75} & \tablenum{5.60} & \tablenum{-0.16} & \tablenum{1.02} & \tablenum{3.44}               & \tablenum{3.74}  \\
HSEsol & \tablenum{5.25} & \tablenum{-4.00} & \tablenum{5.58} & \tablenum{-0.43} & \tablenum{1.06} & \tablenum{3.26}               & \tablenum{3.51}  \\
\textbf{HSE+D3} & \tablenum{5.41} & \tablenum{-0.93} & \tablenum{5.60} & \tablenum{-0.14} & \tablenum{1.03} & \tablenum{3.34}      & \tablenum{3.67}  \\
SCAN0  & \tablenum{5.36} & \tablenum{-1.90} & \tablenum{5.53} & \tablenum{-1.39} & \tablenum{1.03} & \tablenum{3.33}               & \tablenum{5.37}  \\
PBE0   & \tablenum{5.48} & \tablenum{0.37} & \tablenum{5.58} & \tablenum{-0.42} & \tablenum{1.02} & \tablenum{3.44}               & \tablenum{4.44}  \\
Expt.\cite{Beyer_228_1967,Zavabeti_277_2021} & \tablenum{5.46} & --- & \tablenum{5.61} & --- & \tablenum{1.03}  & ---  & \tablenum{3.66}  \\  \bottomrule
\end{tabular}
\end{table}

In Figure~\ref{fig:perfect_TeO2}b, we show the band structure and density of states (DOS) for the bilayer;
the monolayer results are found in Figure S\num{1} in the \textit{Supporting Information}.
A slightly larger band gap of \SI{3.92}{\electronvolt} is observed for the monolayer, aligning with a previous report~\cite{Guo_8397_2018}.
From the DOS, a strong hybridization between the \ce{Te} and \ce{O} states is found at the valence band maximum (VBM) and conduction band minimum (CBM),
which is also related with their spatial distribution shown in Figure~\ref{fig:perfect_TeO2}c.
The VBM, where carrier holes propagate, primarily consists of \ce{Te}-5$sp$ lone pairs and O-$2p$ orbitals.
This delocalization may contribute to the high hole mobility observed in the bilayer ${\beta}$-\ce{TeO2}~\cite{Guo_8397_2018,Zavabeti_277_2021}.

\begin{figure*}[t!]
\centering\includegraphics[width=0.9\linewidth]{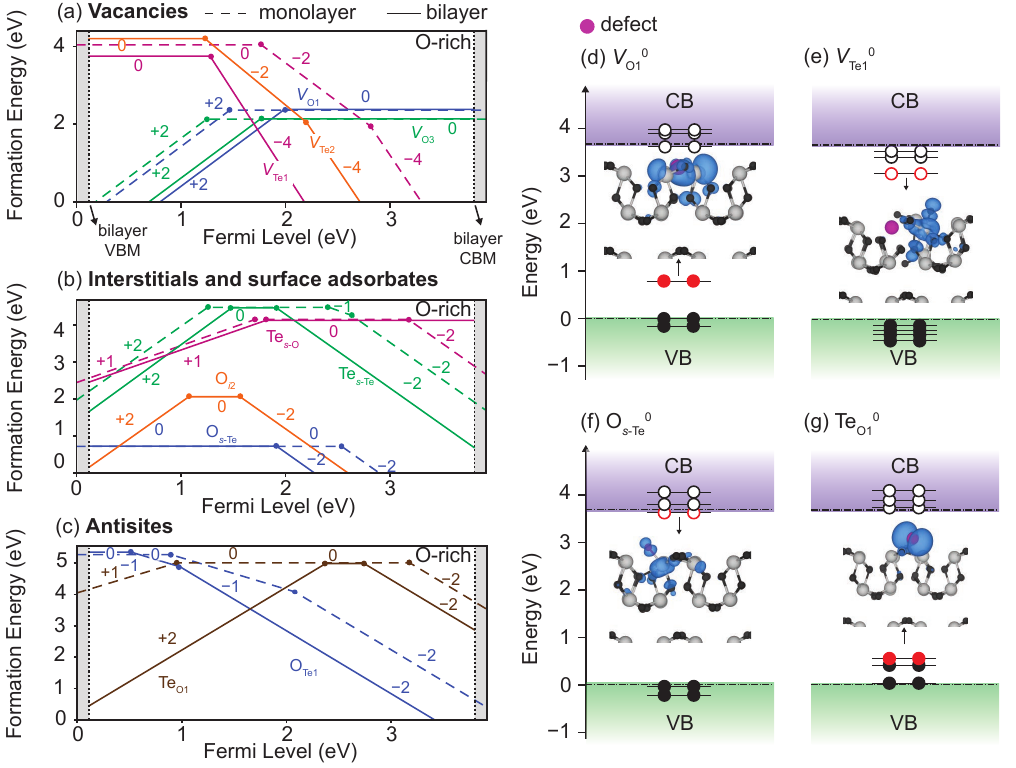}
\caption{(a--c) Formation energies of the (a) vacancies, (b) interstitials and surface adsorbates, and (c) antisites as a function of the Fermi level.
The chemical potentials are set at the O-rich condition.
The dashed and solid lines represent results for the monolayer and bilayer, respectively.
The Fermi levels are referenced to the VBM of the monolayer (see text for details).
    (d--g) Single particle levels of $V_{\text{O1}}$$^0$, $V_{\text{Te1}}$$^0$, O$_{s\text{-Te}}$$^0$, and Te$_{\text{O1}}$$^0$ in the ${\beta}$-\ce{TeO2} bilayer.
    The blue isosurfaces correspond to the squared wave functions labeled with red circles at \SI{0.005}{\elementarycharge\per\cubic{Bohr}}.}
\label{fig:fed_intrinsic}
\end{figure*}

\subsection{Intrinsic defects and unintentional dopants}\label{subsec:intrinsic-defects-and-unintential-dopants}
Next, the point defects in the monolayer and bilayer $\beta$-\ce{TeO2} are explored with the HSE+D3 functional.
In this study, several nonequivalent sites are considered,
as illustrated in Figures~\ref{fig:perfect_TeO2} and S\num{1} in the \textit{Supporting Information}.
The monolayer contains one \ce{Te} site and two nonequivalent \ce{O} sites, located in the outer and inner oxygen layers,
and they are labeled as \ce{Te}\num{1}, \ce{O}\num{1}, and \ce{O}\num{2}, respectively.
In the bilayer, the number of nonequivalent sites doubles because atoms can be located either on the surface side or between the layers.
Consequently, additional sites \ce{Te}\num{2}, \ce{O}\num{3}, and \ce{O}\num{4},
along with a new interstitial site between the two layers (see Figure \ref{fig:perfect_TeO2}a), are introduced.

The formation energies under the \ce{O}-rich condition for the lowest energy defects of each type, \textit{i.e.},
vacancies, interstitials, surface adsorbates, and antisites, are shown in Figure~\ref{fig:fed_intrinsic}a--c.
In this study, the band edges between the monolayer and bilayer are aligned at the vacuum level.
We begin our discussion with the intrinsic vacancies ($V_{\ce{Te}}$ and $V_{\ce{O}}$) in Figure~\ref{fig:fed_intrinsic}a.
Since PBEsol predicts nearly degenerate formation energies for $V_{\ce{O}}$ at all four oxygen sites,
we selectively calculated \ce{O}1 and \ce{O}3 using HSE+D3.
They present very deep donor levels; the $V_{\ce{O}}$ in the bilayer has a ($+2$/$0$) transition level
around \num{1.66} and \SI{1.88}{\electronvolt}, while that in the monolayer is about \num{1.25} and \SI{1.47}{\electronvolt} above their
respective VBMs. The electrons at the in-gap states in the neutral charge states are captured by neighboring \ce{Te} atoms,
as illustrated in Figure~\ref{fig:fed_intrinsic}d.

Even under the \ce{O}-rich (\ce{Te}-poor) conditions,
$V_{\ce{Te}}$$^0$ exhibits relatively high formation energies, around \SI{4}{\electronvolt} in the bilayer.
Here, superscripts in defect notations represent charge states.
This is expected, as forming a \ce{Te} vacancy requires breaking four \ce{Te-O} bonds,
compared to two bonds for an \ce{O} vacancy.
The formation energy of $V_{\ce{Te}1}$$^0$ is \SI{0.5}{\electronvolt} lower than that of $V_{\ce{Te}2}$$^0$,
indicating that $V_{\ce{Te}}$ preferentially forms on the vacuum-facing side.
As shown in Figure~\ref{fig:fed_intrinsic}e, the unoccupied defect states of $V_{\ce{Te}1}$$^0$ are localized in a nearby \ce{TeO5} cluster.
However, these vacancies act as deep acceptors with transition levels at \SI{1.17}{\electronvolt} (\SI{1.77}{\electronvolt}) in the bilayer (monolayer),
making them unlikely to be a main source of hole carriers in 2D ${\beta}$-\ce{TeO2}.

\begin{figure*}[t!]
\centering\includegraphics[width=0.9\linewidth]{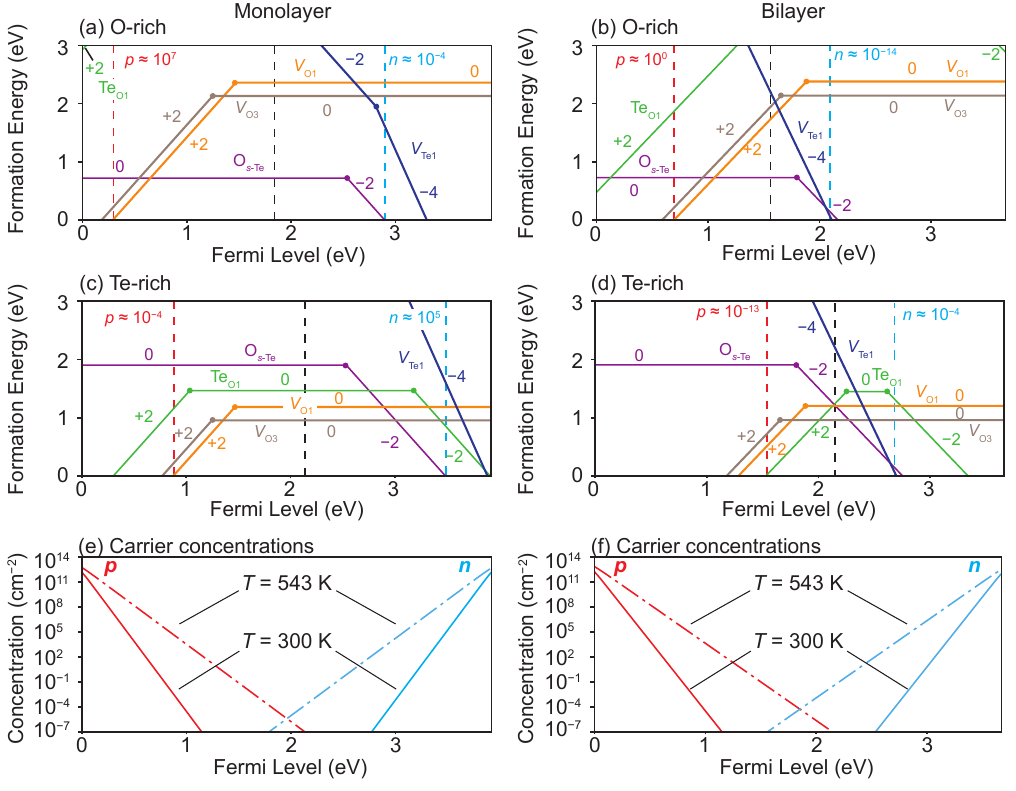}
\caption{(a--d) Same as Fig.~\ref{fig:fed_intrinsic}(a--c) but focusing on dominant native defects
in (a,c) the monolayer and (b, d) the bilayer ${\beta}$-\ce{TeO2}
under (a,b) the O-rich and (c, d) the Te-rich conditions.
    The vertical black dashed lines are the equilibrium Fermi levels at 300~K (see the main text for details).
    The vertical red and blue dashed lines represent the lower and upper limits of the Fermi levels
    at which formation energies of the charged defects are zero, respectively.
    $p$ and $n$ represent the orders of hole and electron concentrations in cm$^{-2}$ at 300~K, respectively.
    (e,f) Carrier concentrations as a function of the Fermi level
    for (e) the monolayer and (f) the bilayer ${\beta}$-\ce{TeO2}.
    The Fermi levels are referenced to each VBM in the diagrams.
}
\label{fig:fed_intrinsic_all}
\end{figure*}

Calculation results of interstitials and surface adsorbates are presented in Figure~\ref{fig:fed_intrinsic}b.
In the bilayer, \ce{O} primarily occupies the interstitial region between the layers ($i2$),
with amphoteric characteristics with transition levels of ($+2$/$0$) and ($0$/$-2$)
at \SI{0.96}{\electronvolt} and \SI{1.57}{\electronvolt}, respectively.
For the surface defects, the \ce{Te} adatoms adsorb preferentially on the \ce{O} on-top site in the neutral charge states,
whereas the \ce{Te} on-top site becomes more stable in charged states,
showing an amphoteric behavior similar to $\ce{O}_{i2}$.
Under the \ce{O}-rich condition, \ce{O} adsorption on top of \ce{Te} (\ce{O}$_{s\text{-}\ce{Te}}$) exhibits a remarkably low formation energy.
In analogy to $V_{\ce{Te}1}$ and $V_{\ce{Te}2}$, \ce{O}$_{s\text{-}\ce{Te}}$ should not be a significant source of hole carriers
due to the ($0$/$-2$) deep acceptor levels, at \SI{2.54}{\electronvolt} in the monolayer
and \SI{1.80}{\electronvolt} in the bilayer, respectively.
As shown in Figure~\ref{fig:fed_intrinsic}f, its hole polaronic states exhibit strong localization
at the \ce{Te} lone pair, as with the Te vacancies.

Due to the large difference in the ionic radii between \ce{O^{2-}} (\SI{1.35}{\AA})
and \ce{Te^{4+}} (\SI{0.66}{\AA})~\cite{Shannon_751_1976}, the
formation of the antisite defects is expected to be energetically unfavorable.
This is indeed observed in Figure~\ref{fig:fed_intrinsic}c,
in which $\ce{O}_{\ce{Te}1}$ and $\ce{Te}_{\ce{O}1}$ show higher formation energies
than the vacancies, interstitials, and surface adsorbates.
However, under the \ce{Te}-rich conditions, the $\ce{Te}_{\ce{O}1}$$^0$ formation energy largely decreases to \SI{1.5}{\electronvolt}.
$\ce{Te}_{\ce{O}1}$ exhibits an amphoteric nature,
with deep ($+2$/$0$) and ($0$/$-2$) transition levels at \SI{2.25}{\electronvolt} and \SI{2.62}{\electronvolt}, respectively.
Because of its donor-like behavior and low formation energy in the $p$-type regime,
$p$-type doping may be inhibited under the \ce{Te}-rich conditions.

In Figure~\ref{fig:fed_intrinsic_all},
the formation energies of the dominant native defects
under the \ce{O}- and \ce{Te}-rich conditions are shown.
Based on the charge neutrality condition~\cite{Kumagai_125202_2014,PRXEnergy.2.043002},
we evaluated the equilibrium Fermi levels ($E_{\text{F}}$),
as well as the defect and carrier concentrations in 2D $\beta$-\ce{TeO2}.
The temperature is initially set to the reported bilayer growth temperature (\SI{543}{\kelvin})~\cite{Zavabeti_277_2021},
and then quenched to \SI{300}{\kelvin}, maintaining defect concentrations
while allowing charge state transitions.

The calculated $E_{\text{F}}$ of the monolayer and bilayer are
located at \SI{1.78}{\electronvolt} and \SI{1.60}{\electronvolt} from the VBMs, respectively, at the \ce{O}-rich conditions,
where $V_{\ce{O}1}$$^{2+}$ and $\ce{O}_{s\text{-}\ce{Te}}$$^{2-}$ determine the $E_{\text{F}}$,
indicating intrinsically insulating characteristics in both the monolayer and bilayer.
For instance, the bilayer presents a hole concentration of \SI{3.4e-15}{\per\square\cm}.
Note that the observed hole concentration determined by the Hall measurement is \SI{1.04e9}{\per\square\cm}~\cite{Zavabeti_277_2021}.
The oxygen adatom is one of the few defects that exists in meaningful concentrations,
\textit{e.g.}, \SI{1e8}{\per\square\cm}.
However, since it remains neutral at the equilibrium Fermi level,
it does not contribute to the $p$-type conductivity.
This suggests that native defects in 2D-\ce{TeO2} are unlikely to be the main source of $p$-type carriers.

The tendencies of the native defects in 2D-\ce{TeO2} are
in line with those of \textit{bulk} $\beta$-\ce{TeO2} we reported recently~\cite{bulk_teo2_2024},
indicating the intrinsic insulating nature in both 2D and 3D $\beta$-\ce{TeO2}.
The notable difference is the $E_{\text{F}}$, which is located at \SI{0.53}{\electronvolt} in bulk $\beta$-\ce{TeO2},
comparatively closer to the VBM. This is because $V_{\ce{Te}}$$^{4-}$ has a lower formation energy in the bulk,
while \ce{O} adsorbates regulate the $E_{\text{F}}$ in 2D $\beta$-\ce{TeO2}.
The dimensionality reduction, however, does not contribute to the hole concentrations, which is the key conclusion of this study.

\begin{figure}[t!]
    \centering\includegraphics[width=0.5\linewidth]{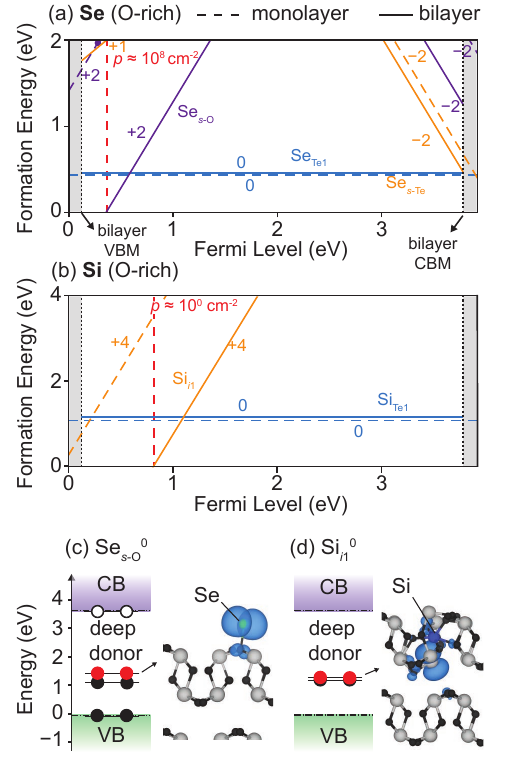}
    \caption{(a,b) Same as Fig.~\ref{fig:fed_intrinsic}(a--c) but for stable substitutional and interstitial defects
    of the \ce{Se} and \ce{Si} impurities under the O-rich conditions.
    Extended formation energy diagrams are displayed in Figure S6.
        (c, d) Same as Fig.~\ref{fig:fed_intrinsic}(d--g) but for (c) $\ce{Se}_{s\text{-}\ce{O}}$$^0$ and (d) $\ce{Si}_{i1}$$^0$.}
    \label{fig:fed_impurities}
\end{figure}

We extend our investigation to unintentional impurities, namely \ce{Se} and \ce{Si}
because their presence was confirmed by XPS elemental mapping of the deposited $\beta$-\ce{TeO2} flakes~\cite{Zavabeti_277_2021}.
Figures~\ref{fig:fed_impurities}a and b show the formation energies of \ce{Se} and \ce{Si} impurities, respectively.
Our calculations show that \ce{Se} predominantly occupies the substitutional \ce{Te} sites ($\ce{Se}_{\ce{Te}1}$),
with a formation energy below \SI{0.5}{\electronvolt}.
Since \ce{Se} and \ce{Te} are in the same group of the periodic table,
it is natural that the substitutional defect remains in a neutral charge state.
In the bilayer, \ce{Se} adsorbed on oxygen site ($\ce{Se}_{s\text{-}\ce{O}}$)
pins the Fermi level at \SI{0.25}{\electronvolt} above the VBM,
which is lower than the pinned level by the oxygen vacancies, resulting in no significant effect.
On the other hand, the \ce{Si} contaminant can efficiently introduce
the inner layer interstitial ($\ce{Si}_{i1}$$^{4+}$) at the $p$-type condition,
and therefore, \ce{Si} should be avoided as much as possible to achieve the $p$-type conductivity.
As illustrated in Figures~\ref{fig:fed_impurities}c and d,
both $\ce{Se}_{s\text{-}\ce{O}}$$^0$ and $\ce{Si}_{i1}$$^0$ defects introduce deep donor states of polaronic nature,
which is related to the donor-type behavior.

Based on our calculations, native defects and unintentional dopants do not contribute to the hole conductivity.
It is noteworthy that there is a notable mismatch of the Fermi levels observed in 2D $\beta$-\ce{TeO2}
using the Hall measurement and optical spectra~\cite{Zavabeti_277_2021}.
The Hall measurement estimates hole concentration of the bilayer to be \SI{1.04e9}{\per\square\cm}.
Based on our calculated carrier concentrations as a function of the Fermi level (Figure~\ref{fig:fed_intrinsic_all}f),
which can be evaluated from the DOS, this concentration corresponds to the Fermi level around \SI{0.2}{\electronvolt} from the VBM.
Meanwhile, the ultraviolet photoelectron spectroscopy (UPS) and STS measurements
for the same sample confirm the location of Fermi level at \SI{0.9}{\electronvolt}.
This large mismatch suggests that hole conduction is not simply mediated by the VBM or its perturbed states.
Instead, we here propose two potential sources for hole conductivity.

The first could be hole propagation through localized defect states.
Both experimental and theoretical studies indicate that
deep localized defect states are common in 2D materials~\cite{Wang_196801_2015,Wang_8_2019,Ma_100304_2023},
probably due to poor charge carrier screening inherent to their reduced dimensionality~\cite{Komsa_031044_2014}.
This would also be related to the impurity conduction in other 2D materials~\cite{Wang_8_2019,Sun_044063_2020,Wang_031002_2022,Bae2024},
thus, a similar scenario could occur in 2D $\beta$-\ce{TeO2}.

The second possible source is the substrate effect.
Charge transfer from the substrate can play a key role, especially in interfaces with 2D materials~\cite{Wang_244701_2020,Wang_1682_2020,Huang_20678_2022}.
We examined the interaction between $\beta$-\ce{TeO2} and the \ce{SiO2} substrate,
considering \ce{Si}-, \ce{Si-H}-, \ce{O}-, and \ce{O-H}-terminated surfaces, which are commonly expected to coexist.
Details of the calculations are provided in the \textit{Supporting Information}.
Our findings reveal no significant charge transfer at the \ce{H}-passivated interfaces.
However, at \ce{Si}-rich interfaces, electron transfer to the nanosheet occurs,
potentially harming its $p$-type conductivity.
On the other hand, the \ce{OH}-termination, which is likely to dominate
due to water vapor in the environment, promotes charge transfer from the nanosheets to the substrate,
potentially enhancing the hole concentration.
The \ce{-SiO} terminations are anticipated to eventually form because
the synthesis temperature of the nanosheets exceeds the dehydroxylation threshold for \ce{SiO2} surfaces
(\ce{-Si(OH)2} $\rightarrow$ \ce{-SiO} + \ce{H2O}), typically around
\SI{200}{\celsius}~\cite{ElRassy_1603_2005,Schmidt_135_2012}. Based on our results, this should further enhance the hole
concentration since a $40\%$ increase in charge transfer is observed for a $8\%$ surface coverage of \ce{-SiO} terminations.
In an extreme scenario, the oxygen-saturated \ce{SiO2} surface, achievable via
dry oxidation, removes \SI{4.06}{\elementarycharge} from the 2D $\beta$-\ce{TeO2} according to the Bader analysis,
completely oxidizing some interface \ce{Te}, which may contribute to the $p$-type conductivity.

\begin{figure*}[t!]
\centering\includegraphics[width=0.9\linewidth]{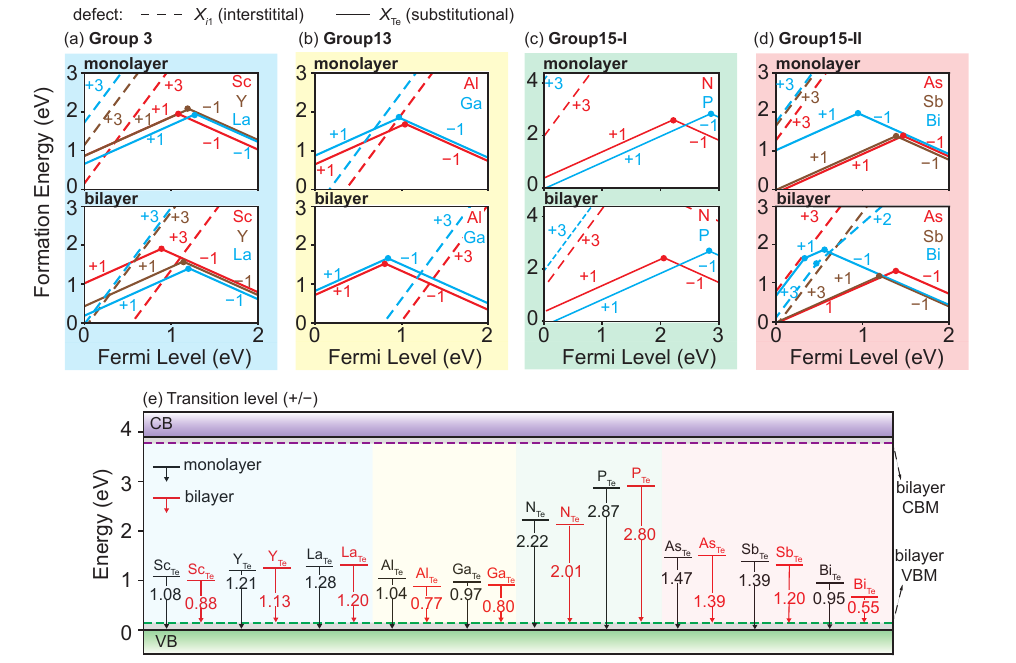}
\caption{
    (a--d) Same as Fig.~\ref{fig:fed_intrinsic}(a--c) but focusing on
    the most stable substitutional and interstitial dopant defects under the \ce{O}-rich conditions.
    (e) The $(+1/-1)$ transition levels of the substituted dopants.}
\label{fig:fed_dopants}
\end{figure*}

\subsection{Screening of acceptor dopants}\label{subsec:screening-of-acceptor-dopants}
We then investigated the existence of shallow acceptor dopants in 2D $\beta$-\ce{TeO2}.
A total of 10 dopants from groups 3 (\ce{Sc}, \ce{Y}, \ce{La}),
13 (\ce{Al}, \ce{Ga}) and 15 (\ce{N}, \ce{P}, \ce{As}, \ce{Sb}, and \ce{Bi}) substituting \ce{Te} sites
were calculated.

The defect formation energies and charge transition levels of the interstitials
and substitutional impurities at the \ce{Te} sites ($X_{\ce{Te}}$) are presented in Figure~\ref{fig:fed_dopants}.
For Group 15 elements, we classify them into 15-I (N, P) and 15-II (As, Sb, Bi)
because these two classes exhibit distinct behaviors from each other.
For the diagrams with extended energy ranges and the other defect types with higher formation energies,
refer to Figures~S7-10 in the \textit{Supporting Information}.
For most dopants, the substitutional sites are the most stable,
followed by the interstitial site $X_{i1}$, which is placed in the layer (see Fig.~\ref{fig:perfect_TeO2}(a)).
The only exception is \ce{N}, which prefers the interlayer space ($i2$ site).

The substitutional $X_{\ce{Te}}$ defects show direct transitions between $-1$ and $+1$ charge states,
known as the negative-$U$ effect.~\cite{}
The effective $U$ value ($U^{\text{eff}}$) between $q-1$, $q$, $q+1$ is evaluated as
\begin{equation}
    U^\text{eff}(q-1/q/q+1) = E_f[D^{q-1}] + E_f[D^{q+1}] - 2E_f[D^{q}],
    \label{eq:equation}
\end{equation}
where $E_f[D^{q}]$ is the formation energy of defect $D$ with charge $q$ (see Sec.~\ref{subsec:point-defect-calculations}).
In Table~\ref{tab:eff_u}, we summarize $U^{\text{eff}}$ for the $X_{\ce{Te}}$ defects,
which range from \SI{-0.15} to \SI{-0.87}{\electronvolt} for the monolayer,
and from $-0.44$ to \SI{-1.54}{\electronvolt} for the bilayer.

We also calculated the self-trapped holes (STH) as a polaron ($q=+1$) and a bipolaron ($q=+2$)
by placing one and two holes in perfect supercells, respectively, and estimated its $U^{\text{eff}}$.
Such STHs are found not to relax to the delocalized states, but their formation energies
are unstable over delocalized hole states (see \textit{Supporting Information}).
$U^{\text{eff}}$ of STHs are \SI{-0.22}{\electronvolt} for the monolayer and \SI{-0.73}{\electronvolt} for the bilayer,
which indicates that the negative $U^{\text{eff}}$ observed for the acceptor dopants is related to the
intrinsic nature of the bipolarons in 2D $\beta$-\ce{TeO2}.
The instability of the single hole originates from the open-shell $4d^{10}5s^{1}$ configuration of $\ce{Te}^{5+}$ ions,
where a hole is captured. Such valence skipping is often observed in lone pair orbitals.
For example, \ce{Bi} in \ce{K}-doped \ce{\textit{X}BiO3}($X$ = \ce{Ba}, \ce{Sr}) prefers
$+3$ and $+5$ oxidation states over the $+4$ state~\cite{Franchini_256402_2009,Nishio_176403_2005,Foyevtsova_121114R_2015}.

Figure~\ref{fig:fed_dopants}e displays the $(+1/-1)$ transition levels for the $p$-type dopants.
Group 15-I dopants exhibit particularly deep acceptor levels in both monolayer and bilayer cases,
making them unsuitable candidates for hole doping.
For the monolayer, although \ce{Ga} and \ce{Bi} are relatively the most promising candidates, their transition levels are still deep.
The acceptor states in the bilayer are generally shallower than those
in the monolayer. Particularly for \ce{Al} and \ce{Bi} dopants, the $(+1/-1)$ transition states shift
down to \SI{0.77}{\electronvolt} and \SI{0.55}{\electronvolt} above the VBM, respectively.
However, unlike the monolayer, most dopants are self-compensated by $X_{i1}$$^{3+}$ interstitial
defects, significantly limiting the hole dopability of the bilayer.
In the bilayer, the defect charge is widely distributed in two layers, leading to a general reduction
in Coulomb energy of $X_{i1}$$^{3+}$  compared to the monolayer,
which reduces the formation energies of the interstitials (see Figure S11 for the plane-averaged charge difference).
Due to its large ionic size that relates to the reduced self-compensation,
\ce{Bi} remains the most promising dopant candidate for the bilayer.

We acknowledge that \ce{Bi} still classifies as a deep acceptor,
compared to the shallow acceptors in conventional 3D semiconductors,
where acceptor transition levels should lie within \SI{0.1}{\electronvolt} above the VBM.
As discussed in Sec.~\ref{subsec:intrinsic-defects-and-unintential-dopants},
the hole conductivity may result from hopping through defect bands\cite{Wang_8_2019,Kang_3615_2020}.
If this holds true, doping $\beta$-\ce{TeO2} with deep impurities at high concentrations could
still enhance $p$-type conductivity, similar to what has been observed in other 2D materials~\cite{Kang_3615_2020}.

\begin{table}
\centering
\caption{
    The effective Hubbard U ($U^{\text{eff}}$), in \SI{}{\electronvolt}, estimated for
the substitutional $X_{\ce{Te}}$ defects in the monolayer and bilayer of ${\beta}$-\ce{TeO2}.
The $(+1/-1)$ transition levels from the VBM in \SI{}{\electronvolt}.}
\label{tab:eff_u}
\begin{tabular}{lcccc} \toprule
        & \multicolumn{2}{c}{monolayer} & \multicolumn{2}{c}{bilayer} \\ \bottomrule
 dopant & $U^{\text{eff}}$ & $(+/-)$ transition & $U^{\text{eff}}$ & $(+1/-1)$ transition   \\ \bottomrule
\ce{Sc} & \tablenum{-0.26} & \tablenum{1.08} & \tablenum{-0.90} & \tablenum{0.88} \\
\ce{Y}  & \tablenum{-0.30} & \tablenum{1.21} & \tablenum{-0.44} & \tablenum{1.13} \\
\ce{La} & \tablenum{-0.34} & \tablenum{1.28} & \tablenum{-0.87} & \tablenum{1.20} \\
\ce{Al} & \tablenum{-0.23} & \tablenum{1.04} & \tablenum{-0.56} & \tablenum{0.77} \\
\ce{Ga} & \tablenum{-0.15} & \tablenum{0.97} & \tablenum{-0.61} & \tablenum{0.80} \\
\ce{N}  & \tablenum{-0.58} & \tablenum{2.22} & \tablenum{-1.54} & \tablenum{2.13} \\
\ce{P}  & \tablenum{-0.87} & \tablenum{2.87} & \tablenum{-1.36} & \tablenum{2.80} \\
\ce{As} & \tablenum{-0.34} & \tablenum{1.47} & \tablenum{-0.56} & \tablenum{1.39} \\
\ce{Sb} & \tablenum{-0.24} & \tablenum{1.39} & \tablenum{-0.72} & \tablenum{1.20} \\
\ce{Bi} & \tablenum{-0.32} & \tablenum{0.95} & \tablenum{-0.57} & \tablenum{0.55} \\
intrinsic & \tablenum{-0.22} & --             & \tablenum{-0.73}  &  -- \\ \bottomrule
\end{tabular}
\end{table}

\section{Conclusions}\label{sec:conclusions}
In this study, we investigated the mechanism of the hole doping observed in 2D $\beta$-\ce{TeO2}~\cite{Zavabeti_277_2021}.
First, we identified that the HSE+D3 hybrid functional accurately reproduces
both the lattice constants and the band gap observed in experiments.
We then applied the HSE+D3 functional to study native defects and
\ce{Se} and \ce{Si} impurities that are unintentionally contaminated in the 2D $\beta$-\ce{TeO2}.
We focused on the monolayer and bilayer $\beta$-\ce{TeO2},
the latter of which has been experimentally synthesized.
Our findings suggest that neither native defects nor unintentional impurities
are likely sources of $p$-type conductivity in 2D $\beta$-\ce{TeO2}.
Instead, we propose two potential mechanisms for hole conductivity:
The first is hole transport through localized defect states,
and the second involves substrate effects.
We demonstrate the latter using 2D $\beta$-\ce{TeO2}/\ce{SiO2} interface models,
revealing that substrates with \ce{OH}- and \ce{O}-rich surfaces may facilitate
hole injection into the nanosheets.

Additionally, we investigated 10 elements from groups 3 (\ce{Sc}, \ce{Y}, \ce{La}), 13 (\ce{Al}, \ce{Ga}),
and 15 (\ce{N}, \ce{P}, \ce{As}, \ce{Sb}, \ce{Bi}) as potential external acceptor dopants.
All dopants are found to exhibit deep acceptor levels with negative $U$ behavior,
with most forming bipolarons bound to neighboring \ce{Te}-${5sp}$ lone pairs.
Among these, Bi shows the shallowest acceptor level in the bilayer,
although the transition levels remain relatively deep.
Such doping could promote hole mobility through hopping
conduction via localized defect states~\cite{Wang_8_2019,Kang_3615_2020}.

We expect these findings provide valuable understanding of the hole conductivity observed in
2D $\beta$-\ce{TeO2} and open new avenues for optimizing its properties
for future electronic applications.

\section{Computational Details}\label{sec:computational-details}
\subsection{First-principles calculations}\label{subsec:first-principles-calculations}
First-principles calculations were based on the spin-polarized density-functional theory
with the projected augmented wave~\cite{Blochl_17953_1994,Kresse_1758_1999}
(PAW) method as implemented in the VASP code~\cite{Kresse_13115_1993}.
For overall calculations, the Heyd-Scuseria-Ernzerhof (HSE06) functional~\cite{Heyd_8207_2003,Krukau_224106_2006}
with the DFT-D3 method of Grimme with zero-damping function (HSE+D3) was employed~\cite{Grimme_154104_2010}.
Information on the PAW potentials is found within the \textit{Supporting Information}.
To reduce the Pulay stress error, a cutoff energy of \SI{520}{\electronvolt}
was used to determine the in-plane lattice parameters of the 2D ${\beta}$-\ce{TeO2}
and calculate the competing phases to build the chemical potential diagram (CPD).
For the remaining total energy calculations, a cutoff energy of \SI{400}{\electronvolt} was used.
The Brillouin zone was sampled using \textit{k}-point densities of \SI{2.5}{\per\angstrom} and \SI{5.0}{\per\angstrom}
for insulating and metallic systems, respectively.
The ${\Gamma}$-point was sampled for the $3{\times}3$ supercells that were used for the point-defect calculations.
For the band structure, the path in the reciprocal space was determined using seekpath~\cite{hinuma2017band}.

The conjugated gradient algorithm was used to obtain the equilibrium atomic structures
with the cutoff force of \SI{0.005}{\electronvolt\per\angstrom}.
For the defect calculations, for which lattice parameters were kept fixed,
the force and total energy convergence criteria were set to
\SI{0.03}{\electronvolt\per\angstrom} and \SI{d-5}{\electronvolt}, respectively.
The dielectric tensors of the monolayer and the bilayer were computed through
the self-consistent response to finite electric fields using the HSE+D3 hybrid functional.
All the VASP input settings were generated using the VISE code (version 0.9.0)~\cite{vise,Kumagai_123803_2021}.

\subsection{Point-defect calculations}\label{subsec:point-defect-calculations}
Point defects in the monolayer and the bilayer ${\beta}$-\ce{TeO2} were modeled using a $3{\times}3$ supercell,
with the slab thickness set to \SI{20}{\AA}.
The initial interstitial sites were identified by locating the charge density minima~\cite{TSUNODA2022111068}.
Initially, the PBEsol functional~\cite{Perdew_136406_2008} was used to screen for the energetically favorable defects.
The atomic positions of the prescreened defects were further relaxed using the HSE+D3 hybrid functional.

The defect formation energy ($E_f$) of defect $D$ in charge state $q$
was calculated as follows~\cite{Freysoldt_253_2014,Kumagai_195205_2014}:
\begin{equation}
E_{f} = \{E[D^{q}]+E_{\text{corr}}[D^{q}]\} - E_{\text{P}} - \sum{n_{i}{\mu}_{i}} + q({\epsilon}_{\text{VBM}} + {\Delta}{\epsilon}_{\text{F}})~,
\label{eq:dfe}
\end{equation}
where $E[D^{q}]$ and $E_{\text{P}}$ are the total energies of the defective and pristine supercells, respectively,
and $E_{\text{corr}}[D^{q}$] is the correction energy for the charged defect $D^{q}$
under periodic boundary conditions (see Sec.~\ref{subsec:corrections-on-defect-formation-energies}).
$n_{i}$ is the number of atoms of element $i$ added ($n_{i} > 0$) or removed ($n_{i} < 0$) to form the defect,
${\mu}_{i}$ is the chemical potential of that element,
${\epsilon}_{\text{VBM}}$ is the energy level of the VBM,
and ${\Delta}{\epsilon}_{\text{F}}$ is the Fermi level with respect to the VBM.

\subsection{Chemical potential diagram}\label{subsec:chemical-potential-diagram}
The chemical potentials are constrained by competing phases under equilibrium conditions.
The chemical potential diagram for the \ce{Te}-\ce{O} binary system,
constructed from PBEsol total energies, is shown in Figure S\num{2}.
The considered competing phases include \ce{Te} ($P3_121$), \ce{TeO2} ($Pbca$),
\ce{Te2O5} ($P12_1\overline{1}$, $C12/m1$),
and molecular \ce{O2} in its triplet state.
The \ce{TeO2} phase is defined by points A (${\mu_{\rm O}}=\SI{-1.68}{\electronvolt}$; ${\mu_{\rm Te}}=\SI{0.00}{\electronvolt}$) and B (${\mu_{\rm O}}=\SI{-0.51}{\electronvolt}$; ${\mu_{\rm Te}}=\SI{-2.36}{\electronvolt}$), representing Te-rich and O-rich conditions, respectively.
A comprehensive list of phases, whose structures were sourced from the Materials Project
Database (MPD)~\cite{Jain_011002_2013},
along with the chemical potentials used in dopant calculations, is available
in the \textit{Supporting Information}.
For defect formation energy calculations based on HSE+D3,
the energy alignment accounts for the energy differences
between the simple substances and the \ce{O2} molecule, using PBEsol and HSE+D3.

\subsection{Corrections on defect formation energies}\label{subsec:corrections-on-defect-formation-energies}
The image-charge correction energy, $E_{\text{corr}}[D^{q}]$, was calculated following
the correction scheme for two-dimensional systems proposed by Noh \textit{et al.}\cite{Noh_205417_2014}
and Komsa \textit{et al.}~\cite{Komsa_031044_2014},
combined with the Sundararaman and Ping approach~\cite{Sundararaman_104109_2017} for isolated Gaussian charges.
Note that the correction method is an extension of the method proposed by Freysoldt et al.~\cite{PhysRevLett.102.016402},
which is known to accurately correct the formation energy of defects in various materials~\cite{PhysRevApplied.6.014009,PhysRevApplied.9.034019,PhysRevApplied.10.011001}.
This was implemented using the PYDEFECT\_2D code~\cite{Kumagai_054106_2024}.
For this correction, the dielectric profiles of 2D $\beta$-\ce{TeO2} were
modeled using a step function.
The thicknesses of the step-like dielectric profiles ($w_z$) were set
to $\SI{5.3}{\AA}$ for the monolayer and $\SI{11.5}{\AA}$ for the bilayer,
with edge smearing $\beta=\SI{0.5}{\AA}$~\cite{Kumagai_054106_2024}.
Additionally, we employed PYDEFECT~\cite{Kumagai_123803_2021} to parse $E_{\text{corr}}[D^{q}]$,
compute defect formation energies, and determine charge transition levels.
Further details are available in the references~\cite{pydefect,Kumagai_054106_2024}.

\subsection{Charge transition level}\label{subsec:charge-transition-level}
A transition level is defined as the Fermi level where the most stable charge state changes.
Using the defect formation energies, the transition level $\epsilon(q/q')$ between charge states $q$ and $q'$,
where ${\Delta}E_{f}[D,q] = {\Delta}E_{f}[D,q']$, is calculated as:
\begin{equation}
 \epsilon(q/q') = \frac{{\Delta}E_{f}[D,q'] - {\Delta}E_{f}[D,q]}{q - q'}.
\end{equation}

When $q > q'$, charge state $q$ is more stable for Fermi levels below $\epsilon(q/q')$, and vice versa.
The Fermi levels are usually referenced relative to the VBM.
The gap between $\epsilon(0/-1)$ and the VBM corresponds to the acceptor level,
and the gap between $\epsilon(+1/0)$ and the CBM represents the donor level.


\begin{suppinfo}
The Supporting Information is available free of charge at https://pubs.acs.org/doi/xxx.
A brief discussion on the Fermi level pinning by defects in 2D $\beta$-\ce{TeO2};
details on the evaluation of the STHs and calculations of $\beta$-\ce{TeO2}/\ce{SiO2} interfaces;
information on the PAW potentials used in this study; the chemical potentials and
competing phases considered to build the chemical potential diagrams;
top and side views of the monolayer and bilayer, together with their respective
band structure and DOS; the chemical potential diagram for the \ce{TeO2}; formation energy diagrams
for the native defects, obtained from PBEsol calculations;
extended formation energy diagrams for the native defects, impurities, and dopants, obtained from HSE+D3 calculations;
partial charge isosurfaces of the selected defect states in the monolayer and bilayer;
charge density difference and their plane average along the $c$ axis for the $\ce{Al}_{i3}^{3+}$ defect
and $\beta$-\ce{TeO2}/\ce{SiO2} interfaces, as well as the interface separation distance and the interface binding energies.
\end{suppinfo}

\section*{Acknowledgments}
The authors thank the support from JSPS KAKENHI Grant Number 22H01755 and 23KF0030
and the E-IMR project at IMR, Tohoku University.
Part of calculations were conducted using MASAMUNE-IMR (Project No. 202312-SCKXX-0408) and ISSP supercomputers.


\providecommand{\latin}[1]{#1}
\makeatletter
\providecommand{\doi}
  {\begingroup\let\do\@makeother\dospecials
  \catcode`\{=1 \catcode`\}=2 \doi@aux}
\providecommand{\doi@aux}[1]{\endgroup\texttt{#1}}
\makeatother
\providecommand*\mcitethebibliography{\thebibliography}
\csname @ifundefined\endcsname{endmcitethebibliography}
  {\let\endmcitethebibliography\endthebibliography}{}

\end{document}